\documentclass[
aip, rsi, amsmath, amssymb, reprint, altaffillsymbol
]{revtex4-1}

\usepackage{graphicx}
\usepackage{dcolumn}
\usepackage{bm}

\usepackage{siunitx}
\usepackage{comment}
\usepackage{braket}

\usepackage[utf8]{inputenc}
\usepackage[T1]{fontenc}
\usepackage{mathptmx}
\usepackage{etoolbox}
\usepackage{xcolor}
\usepackage{csquotes}
\usepackage{upgreek}

\def\adjust@abstractwidth{%
 \parindent1em\relax
 \advance\leftskip.5in\relax
 \@totalleftmargin\leftskip
 \@afterheading\@afterindentfalse
}%
\def\frontmatter@abstractheading{}%
\def\frontmatter@abstractfont{%
 \adjust@abstractwidth
}%

\makeatletter
\def\@email#1#2{%
 \endgroup
 \patchcmd{\titleblock@produce}
  {\frontmatter@RRAPformat}
  {\frontmatter@RRAPformat{\produce@RRAP{*#1\href{mailto:#2}{#2}}}\frontmatter@RRAPformat}
  {}{}
}%
\makeatother
\begin{document}

\title[Active Stabilization of Laser Diode Injection]{Active Stabilization of Laser Diode Injection Using a Polarization-Spectroscopy Technique}

\author{Luka Milanovic} 
\email[]{mluka@ethz.ch}
\altaffiliation{These authors contributed equally.}
\affiliation{ETH Z\"urich - PSI Quantum Computing Hub, 5232 Villigen PSI, Switzerland}
\affiliation{Institute of Quantum Electronics, ETH Z\"urich, Otto-Stern-Weg 1, 8093 Zurich, Switzerland}
\author{Greg Ferrero} 
\altaffiliation{These authors contributed equally.}
\affiliation{ETH Z\"urich - PSI Quantum Computing Hub, 5232 Villigen PSI, Switzerland}
\affiliation{Institute of Quantum Electronics, ETH Z\"urich, Otto-Stern-Weg 1, 8093 Zurich, Switzerland}
\author{Robin Oswald} 
\affiliation{ETH Z\"urich - PSI Quantum Computing Hub, 5232 Villigen PSI, Switzerland}
\affiliation{TEM Messtechnik GmbH, 30559 Hanover, Germany}
\author{Thomas Kinder} 
\affiliation{TEM Messtechnik GmbH, 30559 Hanover, Germany}
\author{Julian Schmidt} 
\affiliation{ETH Z\"urich - PSI Quantum Computing Hub, 5232 Villigen PSI, Switzerland}
\author{Cornelius Hempel} 
\affiliation{ETH Z\"urich - PSI Quantum Computing Hub, 5232 Villigen PSI, Switzerland}
\affiliation{Institute of Quantum Electronics, ETH Z\"urich, Otto-Stern-Weg 1, 8093 Zurich, Switzerland}
\date{\today}

\begin{abstract}
	\rightskip.5in
Laser diode injection-locking is a commonly used method to amplify laser light, while preserving its spectral properties. Fluctuations in the environmental conditions can cause injection-locking to fail, especially when operating with low seed powers or with a swept seed frequency. We present a method inspired by the H\"ansch-Couillaud scheme to monitor and actively stabilize the conditions required for injection-locking a laser diode. Using only a few optical components, our scheme can run continuously in the background and is modulation-free. We demonstrate its efficacy by showing its robustness to large fluctuations in diode temperature, seed frequency and power, effectively extending the reliable operating range and stability over time.
\end{abstract}

\maketitle

\section{Introduction}

Laser diode injection-locking is often used to obtain higher optical powers from a low power laser source while retaining its spectral properties. Injection-locking is particularly useful in high power applications, since it can achieve the required high gains more easily than optical amplifiers. Diode injection-locking works passively and requires the incoming light (the seed) to be well-matched to the injected laser diode (ILD) in terms of its mode profile, polarization and frequency. For successful operation, the temperature and the drive current of the injected diode typically have to be carefully tuned to a suitable operating point. However, drifts in the environment or changes in the seed light frequency or power can cause the injection-lock to fail, thus making the technique less resilient than optical amplifiers.

In laboratory settings, injection-setups often rely on passive stability of the operating conditions paired with occasional manual re-adjustments. There are also automatic stabilization techniques at various levels of sophistication. Saxberg et al.\cite{Saxberg2016} diagnose the injection state via a scanning Fabry-Pérot cavity and re-adjust it using a microcontroller instead of a human operator. Similarly, there are also a number of schemes exploiting various physical properties of the light from the ILD such as its spectrum\cite{ChenEtAl2021,RatkoceriBatagelj2021}, its phase shift relative to the seed light\cite{BordonalliEtAl1999}, its polarization extinction ratio\cite{NiederriterEtAl2021}, or its intensity\cite{ShafferEtAl2008,kiesel2024}. The main drawback of these approaches is that they typically only react once the measured property of the light from the ILD has already deteriorated noticeably. Furthermore, they are often restricted to a specific operating point where the physical features exploited in the measurement-scheme are present, but this may not coincide with where one would ideally like to operate the diode. Both these issues can be addressed using more general approaches based on commonly used laser stabilization techniques that measure the detuning between a laser and a cavity such as Pound-Drever-Hall-locking\cite{Drever1983}, tilt-locking\cite{ShaddockEtAl1999}, squash-locking\cite{DioricoEtAl2024} and H\"ansch-Couillaud-locking\cite{Hansch1980} (HC), which have all been adapted to the case of stabilizing laser injection setups \cite{KimEtAl2015,MuellerUeda1998,OttawayEtAl2001,MishraEtAl2023}.

Here, we present a further adaptation of the HC technique to stabilize an injected laser diode. We derive an error signal that is proportional to the detuning between the ILD and the seed light frequency, based on measuring the detuning-dependent phase shift between light from the seed and the ILD by encoding it into an elliptical polarization. We achieve stable long-term operation over a wide range of environmental and seed light conditions. The main advantages of our approach are that it is modulation-free and can run continuously in the background without disturbing the properties of the light of the ILD, that it reacts well before the injection-lock is lost, that it only requires a few comparably cheap components, and that it is simple to build and align. During the literature review conducted for this manuscript, we found that a similar technique has previously been used to stabilize an injected ring laser\cite{MuellerUeda1998}. Our work instead is an adaptation to the specific case of a laser diode, which entails additional effects due to the amplitude-phase coupling inside of the diode and accordingly different practical considerations.

The remainder of this paper is structured as follows: In Sec. \ref{sec:BG} we first summarize the relevant theory behind laser diode injection-locking and the classic HC error signal technique, and then describe our adaptation of the latter. In Sec. \ref{sec:experiment} we first describe our experimental setup and then show and compare our experimental error signal with the signals obtained from other commonly used diagnostic tools. Next, we close the loop and demonstrate the much extended operating range enabled by our technique. In Sec. \ref{sec:discussion} we conclude by summarizing the advantages and disadvantages of our stabilization scheme and discuss its potential applications.

\newpage

\section{Background}\label{sec:BG}
\subsection{Laser diode injection-locking}\label{sec:diode_inj_theory}

The theory behind optical injection-locking of semiconductor laser diodes is well-developed\cite{Lang1982,MogensenEtAl1985,MurakamiEtAl2003}. We quickly summarize the key aspects relevant for our work following two recent review papers\cite{LauEtAl2009,LiuSlavik2020}. A free-running laser diode with frequency $\omega_\mathrm{d}$ is \enquote{injected} and forced to emit at the frequency $\omega_\mathrm{s}$ of a seed laser, if the detuning $\Delta\omega = \omega_\mathrm{s} - \omega_\mathrm{d}$ is within the range
\begin{equation}
	-  \kappa \sqrt{\frac{P^\prime_\mathrm{s}}{P_\mathrm{d}}} \sqrt{1+\alpha^2} \equiv \Delta\omega_\mathrm{min} < \Delta\omega < \Delta\omega_\mathrm{max} \equiv \kappa \sqrt{\frac{P^\prime_\mathrm{s}}{P_\mathrm{d}}}, \label{eq:lock_range}
\end{equation}
where $\kappa$ is the rate at which the injected seed photons enter into the laser diode, $P^\prime_\mathrm{s}$ is the effective optical power of the injected seed light taking spatial mode and polarization mismatches into account, $P_\mathrm{d}$ the optical power of the light emitted by the injected laser diode, and $\alpha$ the laser linewidth enhancement factor of the laser diode capturing the amplitude-phase coupling of the gain medium.

Staying within the bounds given by Eq.~\eqref{eq:lock_range} becomes harder as $P^\prime_\mathrm{s}$ is reduced, $P_\mathrm{d}$ is increased, or if the seed frequency $\omega_\mathrm{s}$ varies over time. 
The injected laser diode emits light with a phase shift $\phi$ relative to the seed, which depends on the detuning $\Delta\omega$ following
\begin{equation}
	\phi(\Delta\omega) = \sin^{-1}\left( - \frac{\Delta\omega}{ \kappa  \sqrt{\frac{P^\prime_\mathrm{s}}{P_\mathrm{d}}} \sqrt{1+\alpha^2} } \right) - \tan^{-1}(\alpha). \label{eq:lock_phase}
\end{equation}

Typically, the operating parameters are initially carefully set manually and the detuning then starts drifting with environmental conditions, eventually crossing the bounds and thus requiring re-adjustments.
Moreover, in some applications, where specific amplitude and frequency modulation characteristics of the ILD are desirable\cite{LauEtAl2009}, or the output phase of the ILD should be exactly controlled, tight control over the detuning $\Delta\omega$ is required. We will show how this can be achieved by adapting the HC polarization spectroscopy technique.

\subsection{Classic H\"ansch-Couillaud technique}\label{sec:theory_classic_HC}
\begin{figure}
	\centering
	\includegraphics[scale=0.9]{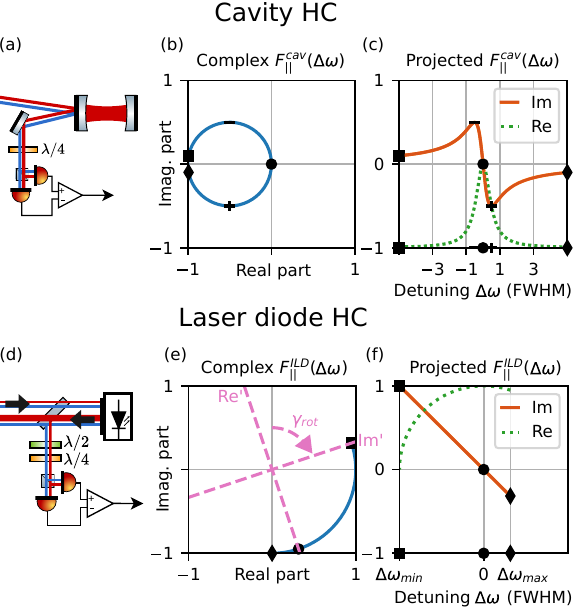}
	\caption{\textbf{Comparison of the classic cavity-based H\"ansch-Couillaud (HC) technique with our adapted version for active stabilization of diode injection-locks.} (a) Basic setup used in classic HC. Only one polarization (drawn in red) can build up in the cavity, and the orthogonal one (drawn in blue) is promptly reflected. The polarization of the back-reflected laser beam is analyzed using a quarter-wave plate and a polarizing beam splitter (PBS) followed by two photodiodes (PDs) whose signals are subtracted to generate the error signal. (b) Complex reflection coefficient $F^\mathrm{cav}_{||}(\Delta\omega)$ of a cavity from Eq.~(\ref{eq:F_cav}) for $R \approx 1$. (c) Projections of the complex reflection coefficient along the imaginary or real axis. The former is the HC error signal.
    (d) Basic setup used for stabilization of diode injection. For clarity the incoming and outgoing beam are drawn separately, but in practice they follow the same path. (e) Complex reflection coefficient $F^\mathrm{ILD}_{||}(\Delta\omega)$ of the laser diode. (f) Projections of the complex reflection coefficient resulting in the error signal (imaginary part) or an auxiliary lock status monitoring signal (real part) depending on the settings of the waveplates before the PBS. The signals in (e) and (f) were calculated for $\alpha=3$.}
	\label{fig:HC_theory}
\end{figure}

The HC polarization spectroscopy technique\cite{Hansch1980} can be used to measure the detuning between a cavity and a laser. A conceptual illustration of the setup typically employed is shown in FIG.~\ref{fig:HC_theory}(a). The back-reflection of a linearly polarized laser beam impinging on a cavity differs strongly for the two orthogonal linear polarizations components due to a polarizing intra-cavity element -- often a non-linear crystal present in the cavity. 

The polarization component encountering strong intra-cavity losses serves as the reference beam since it gets promptly reflected by the input mirror regardless of detuning. The orthogonal polarization component serves as probe beam as it enters the cavity and partially leaks back out with an amplitude and phase given by the detuning-dependent, complex reflection coefficient\cite{Black2001} 

\begin{equation}
	F^\mathrm{cav}_{||}(\Delta\omega) = \frac{R \left( \exp\left( i\frac{\Delta\omega}{\nu_\mathrm{fsr}}\right)-1 \right)}{1- R^2\exp\left(i\frac{\Delta\omega}{\nu_\mathrm{fsr}}\right)}, \label{eq:F_cav}
\end{equation}

where $R$ is the reflection coefficient of the mirrors determining the cavity linewidth and $\nu_\mathrm{fsr}$ is the free-spectral-range of the cavity. The handedness of the combined, now elliptically polarized, return beam hence directly depends on the detuning and can be turned into a suitable error signal.

In the complex plane $F^\mathrm{cav}_{||}(\Delta\omega)$ has the shape of a circle, as illustrated in FIG.~\ref{fig:HC_theory}(b) and shown by Black\cite{Black2001}. As shown in the appendix, the HC technique effectively extracts the imaginary part and produces the error signal shown in FIG.~\ref{fig:HC_theory}(c). As the detuning goes from negative to positive infinity, we traverse the circle clockwise and thus obtain the classic HC error signal featuring two inflection points.

\subsection{Adapted H\"ansch-Couillaud technique for laser diode injection}\label{sec:theory_inj_HC}

A conceptual illustration of the key elements relevant in our adaptation of the HC technique is shown in FIG.~\ref{fig:HC_theory}(d). Light from the seed featuring two orthogonal polarizations hit the entrance facet of the laser diode. One polarization is again promptly reflected and thus serves as the reference beam. The other polarization enters the laser diode and seeds it, ideally leading to successful injection-locking. Part of the light inside of the laser diode is then coupled out, leading to an outgoing beam that is typically much stronger than the incoming seed light.

The light emitted by the ILD exhibits the detuning-dependent phase shift $\phi(\Delta\omega)$ from equation (\ref{eq:lock_phase}), which is imprinted onto the total polarization and can be analyzed with a similar scheme as described in Section~\ref{sec:theory_classic_HC}. We again consider the complex plane representation in FIG.~\ref{fig:HC_theory}(e), noting that 
\begin{equation}
    F^\mathrm{ILD}_{||}(\Delta\omega) \approx e^{i \phi(\Delta\omega)} \label{eq:F_inj}
\end{equation}
in case the condition from Eq.~\eqref{eq:lock_range} is satisfied. In contrast with FIG.~\ref{fig:HC_theory}(b), the allowed range for $\phi(\Delta\omega)$ given in Eq.~\eqref{eq:lock_phase} now only extends from $-\pi/2$ to $+\cot^{-1}(\alpha)$, thus reducing the shape from a full circle to an arc. In addition, the arc is now centered around the origin as the amplitude of the light coming out of the ILD primarily depends on the diode current and is to good approximation independent of the detuning and the seed light strength. Because the light from the ILD is much stronger than that of the seed, it is advisable to use a polarization-dependent pick-off to roughly equalize the amplitude of the two polarizations to be analyzed.

We then obtain a suitable error by extracting the right projection of $F^\mathrm{ILD}_{||}(\Delta\omega)$ in the complex plane using the usual arrangement. As illustrated in FIG.~\ref{fig:HC_theory}(e) we want to extract the imaginary part in a coordinate system rotated by an angle $\gamma_\mathrm{rot}$, which we can do thanks to the additional $\lambda/2$ waveplate. 

As derived in the appendix, the error signal is proportional to the electric field amplitudes of the seed and of the diode output, $\sin(\theta)$ with $\theta$ denoting the angle between the linear polarization emitted by the ILD, relative to the seed and the detuning $\Delta\omega$. In contrast, the allowed range for $\Delta\omega$ as described by Eq.~\eqref{eq:lock_range} is proportional to $\cos(\theta)$.

Additionally, one can simultaneously also extract the other quadrature of $F^\mathrm{ILD}_{||}(\Delta\omega)$ to serve as a lock-status monitor-signal using a second analysis branch. This could be useful for automatic re-locking routines in case the injection-lock is lost.

\section{Experiment}\label{sec:experiment}

\subsection{Setup}\label{sec:exp_setup}
\begin{figure}
    \centering
    \includegraphics[scale=1]{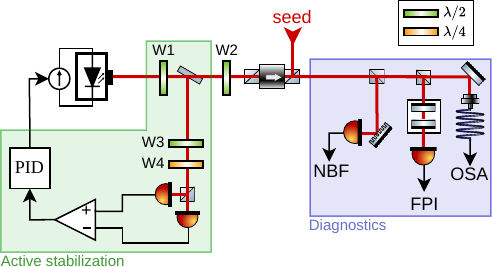}
   \caption{\textbf{Injection setup with H\"ansch-Couillaud-like (HC) error signal generation.} A Faraday isolator is used to introduce light into the injected laser diode's beampath. The waveplates W1 to W4 are used to fully control the polarization in the system. Marked in green are the components that are necessary for generating the HC error signal and active stabilization via feedback onto the diode current. For additional diagnostics, the light is also sent to a narrow-band filter (NBF), a Fabry-Pérot interferometer (FPI), and an optical spectrum analyzer (OSA). The pick-offs for these paths are non-polarizing beamsplitters. For clarity, we omitted a spherical and a cylindrical telescope we use to match the beam profile of the seed to that of the ILD.}
    \label{fig:exp_setup}
\end{figure}

To experimentally demonstrate the scheme outlined above, a standard injection-lock setup~\cite{Greg2024} was built and modified by adding components for active stabilization and additional diagnostics (FIG.~\ref{fig:exp_setup}). A 729 nm laser diode (Thorlabs HL7302MG) is housed in a mount with integrated thermo-electric element (Thorlabs LDM56/M) and its temperature and current are governed by a digital laser diode driver (Koheron CTL200-0), which in turn is controlled via a PID controller (TEM Messtechnik LaseLock). The seed light from a titanium-sapphire laser (MSquared SolsTiS) is delivered to the setup via an optical fiber. This type of source allows for a continuous sweep of the laser wavelength over a large frequency range to probe the robustness of the injection lock. A Faraday isolator is used to isolate the ILD from any back-reflections, as well as to inject the seed light via its third port. A half-wave plate (W2) is used to match the polarization exiting the Faraday isolator to that of the laser diode. \par

The section marked in blue in FIG.~\ref{fig:exp_setup} shows three commonly used tools for diagnosing the state of injection-locked lasers, namely a narrowband filter (Thorlabs FBH730-10, with the incidence angle chosen to optimally separate the free-running wavelength of the ILD from the seed laser wavelength), an optical spectrum analyser (Ocean Optics HR2 VIS-NIR), and a scanning Fabry-Pérot cavity (Thorlabs SA30-73). These tools are not required for normal operation of our scheme and are only used as additional diagnostic tools for characterization purposes. \par

For active stabilization, a small portion of the diode's output light is sampled by a glass plate and analyzed using the HC scheme. As previously mentioned, the light should be sampled using a polarization-dependent pick-off chosen to roughly equalize the power of both polarizations after the pick-off. Here, we achieve this by orienting the glass plate near the Brewster's angle, combined with half-wave plate W1 before it for additional flexibility. The polarization of the sampled light is controlled by tuning waveplates W3 and W4. These degrees of freedom are necessary to correct for constant phase shifts between the two polarizations due to effects in the laser diode and the pick-off. An error signal proportional to the detuning $\Delta\omega$ is then obtained by subtracting the signals of the two photodiodes placed after the PBS.\par

To tune the waveplates for generating a useful error signal, we found the following iterative optimization procedure to be reliable: First the ILD was left completely free running (no seed light) and W1 was tuned to minimize reflection off the pick-off. Next the seed light was turned on and W2 rotated to achieve a large injection range by nearly matching the polarization of the seed light to that of the ILD. A deliberate, small polarization mismatch, of typically 5 to 10 degrees, is necessary to allow for the generation of the error signal. Interpretation of the shape of the error signal is not straightforward and will be discussed Sec. \ref{sec:err_signal_exp}. The error signal's shape can be optimized by tuning W3 and W4. We found a beam-walking-like approach to be effective, where one of the waveplates is slowly rotated, while the other one is used to compensate the DC offset. The shape of the error signal affects the reliability of the feedback loop, which we found to be optimal when maximizing the size of the linear regime around an optimal injection point.\par

\subsection{Error signal}\label{sec:err_signal_exp}
\begin{figure}
    \centering
    \includegraphics[scale=1]{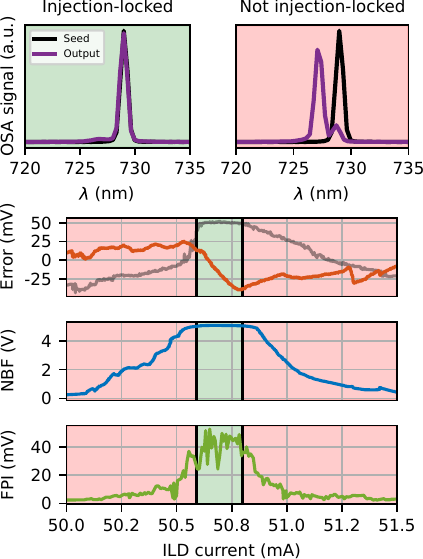}
   \caption{\textbf{HC error signal and further diagnostics as a function of injected laser diode (ILD) current.} In regions where the diode is injection-locked (green area), we observe a linear error signal (red trace) and the optional lock status monitor signal (gray trace) is high. When the detuning between the free running ILD and the seed becomes too large, the injection-lock is lost and the diode lases at a different wavelength as observed on an  optical spectrum analyzer (OSA). In this red region, the error signal primarily depends on the properties of the diode and is not very reliable and thus also not very useful.
   Additionally, we show signals from two other commonly used diagnotic tools, namely the reflection from a narrow-band filter (NBF) and the peak transmission of a scanning Fabry-Pérot interferometer (FPI). To record the FPI trace the current was swept slowly, while the voltage of the FPI was swept quickly. The jaggedness of the FPI line is an artifact from our data acquisition not being synchronized with the sweep.}
    \label{fig:error_map}
\end{figure}

The error signal produced by this scheme is shown in FIG.~\ref{fig:error_map}. In the region where there is injection-locking, the error signal is approximately linear, as expected from Fig.~\ref{fig:HC_theory}(f). Using the waveplates W3 and W4 it is possible to measure different projections of the complex reflection coefficient and thus completely change its shape. FIG.~\ref{fig:error_map} shows the resulting error signal for a good configuration of the waveplates in red. Similarly, a suitable lock status monitor signal can be obtained with different waveplates settings, as shown in gray. When the diode is not injection-locked (red region), the signals primarily depend on the behavior of the laser diode because multiple wavelengths of the light then coexist simultaneously.\par

\subsection{Active stabilization}\label{sec:lock_performance}
Having obtained a suitable error signal, we now implement an active stabilization that slowly adjusts the ILD current using a PID controller. This ensures that the detuning $\Delta\omega$ stays within the allowed bounds, thus enabling reliable operation of the injection-lock even in the face of significant fluctuations of the operating conditions such as changes in the ambient temperature, the frequency of the seed light, or the seed power.
\begin{figure*}
    \centering
    \includegraphics[scale=1]{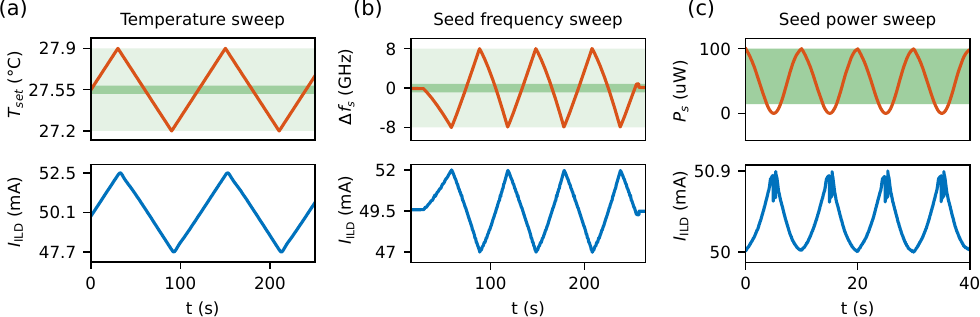}
   \caption{\textbf{Extended operation region achieved via active stabilization.} The feedback loop keeps the laser diode injection-locked by adjusting its current (blue) as various parameters (red) are varied: (a) The temperature is slowly swept by 0.7~$^\circ$C over 30~s requiring current adjustments of 5~mA to compensate for. (b) The seed light frequency is varied by 16~GHz again requiring current adjustments of 5~mA to compensate for. (c) The seed light is varied from 100~$\mu$W down to 0~$\mu$W requiring current adjustments of ~0.9~mA. Measurements (a) and (b) were performed with seed power $P_{\mathrm{s}} = 10$~$\mu$W and average operating current $I_{\mathrm{ILD}}$ = 50~mA resulting in an output power of 22~mW. The dark green areas represent the typical range where the diode stays injection-locked even without active feedback. The light green area represents the extended range where we stay injection-locked due to the feedback loop. Here this range is primarily limited by the range of the current controller rather than our scheme.
   In (c) the injection is lost below a seed power of 5~$\mu$W leading to temporary jumps in $I_{\mathrm{ILD}}$. When enough seed power is re-applied, the feedback loop re-locks without requiring manual intervention.}
    \label{fig:sweeps}
\end{figure*}
To demonstrate the effectiveness of the stabilization we intentionally vary environmental parameters and measure the ranges where the injection-lock is maintained with and without active stabilization. For these measurements, we operate our laser diode at 27~$^\circ$C and 50~mA drive current resulting in 22~mW of optical output power. We use 10~$\upmu$W of total seed light (measured directly before the ILD). 

We first study the sensitivity to changes of the ILD temperature  by varying the setpoint of its temperature controller, see FIG.~\ref{fig:sweeps}(a). Without active stabilization a change of approximately 0.05~$^\circ$C leads to a loss of the injection-lock. With active stabilization the injection current is continuously adjusted to compensate for the temperature changes and the diode stays injection-locked for changes of up to 0.7 $^\circ$C. We stress that this limit is solely due to the restricted current adjustment range of our diode current controller ($\pm$2.5~mA) rather than a fundamental limitation of our scheme. We note, however, that these significant changes in the drive current also lead to output power changes of 6~mW.

Next, we study the effect of varying the seed light frequency as shown in FIG.~\ref{fig:sweeps}(b). Without active stabilization, the seed frequency can be swept by about 1 GHz without losing the injection-lock. With the active stabilization enabled, this value increases to 16 GHz and is again limited by the adjustment range of the current controller. Note that we scanned the seed frequency quite slowly in order to stay well within the bandwidth of our slow stabilization loop. If desired, one can make the stabilization loop much faster and we have also reached bandwidths of about 10~kHz without much effort.

Lastly, we vary the power of the seed light as shown in FIG.~\ref{fig:sweeps}(c). This not only shifts the optimal diode drive current due to slight changes in the diode temperature, but also drastically changes the injection range and the shape of the error signal. We observe that below seed powers of approximately 5~$\mu$W, it is not possible to achieve a reliable injection-lock regardless whether active stabilization is enabled or not. This is also visible in the ILD current fluctuations around this point. As the seed power is increased again, the lock is re-acquired automatically. This shows that the active feedback loop is robust to seed power variations, but does not decrease the minimum seed power required for successful injection-locking. The active stabilization does however, make it much more practical to operate injection-lock setups with little seed power by compensating for disturbances in the other parameters that might cause injection to otherwise fail more quickly.

\section{Discussion}\label{sec:discussion}
Our main motivation for the implementation of the presented stabilization scheme was to cancel slow environmental drifts, in order to enable reliable long-term operation of the injection-lock. For this reason, we intentionally made the feedback loop slow. In practice, this specific application might benefit even more from feedback onto the setpoint of the ILD temperature controller, instead of the ILD current. This would have two main benefits: First, the optical output power of the ILD no longer changes drastically. Second, and more subtly, there is no longer any risk of imprinting electrical noise that may be present on the control signals onto the light from the ILD. If, on the other hand, a fast active stabilization is desired, say due to fast sweeps of the seed frequency, the current control port is of course more suitable because it allows for much faster adjustments.

In terms of features, our HC-inspired scheme is roughly comparable to adaptations of other standard techniques from laser stabilization such as PDH-\cite{KimEtAl2015,MuellerUeda1998}, tilt-\cite{OttawayEtAl2001} and squash-locking\cite{MishraEtAl2023}. For the particular case of laser diode injection-locking, we find that our scheme is particularly well-suited due to a  practical consideration: Compared to PDH, our scheme does not require and imprint any modulation onto the light or the addition of expensive active optical devices such as an EOM. From a conceptual level, squash-locking is most similar to our scheme, but requires careful alignment of a quadrant photodiode because the error signal is obtained from spatial interference of modes as opposed to orthogonal polarization as in our scheme. \par

In summary, we have presented a scheme to actively stabilize the injection-lock of laser diodes and investigated its performance under the effect of varying conditions. Setting up the error signal generation scheme only requires a few standard optical components. We found that active stabilization makes the laser diode injection significantly more resilient against changes in temperature, seed power, and frequency. For otherwise identical operating conditions, our scheme can reduce the frequency of necessary user intervention and increase up-time of the laser system. This can be critical in experiments relying on large numbers of lasers such as quantum information processing with trapped ions and atoms. Even single ILDs may benefit from this method if challenging wavelengths or low seed powers are involved. We envision, e.g., applications in quantum control of atoms and molecules, which may require amplification of laser light transmitted through a high-finesse cavity or higher-order sidebands of an EOM for widely tunable, phase-stable lasers.

\section*{Author contributions}
TK originally came up with the idea and successfully implemented it in a setup used to amplify a 633~nm HeNe laser already back in 2012 (unpublished). The current setup was constructed by LM and GF with input from all authors. LM, GF, and RO performed the measurements and derived the theory. JS and CH supervised the project. All authors contributed to discussing the results and writing the manuscript.

\section*{Acknowledgments}
We thank Jackson Ang'ong'a for stimulating discussions about stabilizing laser diode injection-locks which ultimately led to this project, Edgar Brucke and Yingying Cui for help with setting up the seed laser, and Moritz Fontboté-Schmidt for helpful discussions about injection-locking laser diodes. This work was funded through the ETH Zurich — PSI Quantum Computing Hub. 

\section*{Conflicts of interests}
The authors have no conflicts of interests to declare. RO and TK are employees of TEM Messtechnik, whose LaseLock PID controller was used for measurements. However, this specific device is not required for the scheme presented above.

\section*{Data availability}
The data that supports the findings in this study is available from the corresponding author upon reasonable request.
\newpage
\section*{References}
\bibliography{hc_inj_lock_refs}

\newpage

\section*{Appendix}\label{sec:appendix}
\subsection{Derivation of the error signals}\label{sec:app:theory}
Here we derive the error signal for both classic cavity-based HC and our adapted version for laser diode injection-locking.

\subsubsection{Cavity-based H\"ansch-Couillaud}\label{sec:app:theory:HC}
In cavity-based HC, the light before the cavity is linearly polarized at an angle $\theta$ relative to pure horizontal polarization, i.e.
\begin{equation}
	E^\mathrm{(i)} = E_0 \left( \cos(\theta)\hat{H} + \sin(\theta)\hat{V} \right) \label{eq:app:E_i_HC}
\end{equation}
with $E_0$ denoting the electric field amplitude and $\hat{H}$ and $\hat{V}$ referring to the corresponding polarization quadratures. The back-reflection from the cavity is then given by 
\begin{align}
	E^\mathrm{(r)} &= E_0 \left( \cos(\theta) F^\mathrm{cav}_{||} \hat{H} + \sin(\theta) F^\mathrm{cav}_{\perp} \hat{V} \right) \\ \label{eq:app:E_r_HC}
	             &\equiv E^\mathrm{(r)}_{||} \hat{H} + E^\mathrm{(r)}_{\perp} \hat{V}
\end{align}
where $F^\mathrm{cav}_{||}$ and $F^\mathrm{cav}_{\perp}$ denote the (detuning-dependent) reflection coefficients of the cavity for the two polarizations. We assume that the horizontal polarization probes into the cavity with $F^\mathrm{cav}_{||}$ from Eq.~(\ref{eq:F_cav}), whereas the vertical polarization gets reflected straight away by the entrance mirror with reflectivity $R_1$, i.e. $ F^\mathrm{cav}_{\perp} = \sqrt{R_1}$.

The reflected light is then analyzed by the arrangement consisting of the $\lambda/4$ waveplate, the polarizing beam splitter (PBS) and the two photodiodes (PDs). The Jones matrix of the quarter-wave plate (QWP) at 45 degrees before the PBS is given by
\begin{equation}
	\mathrm{QWP(45^\circ)} = \frac{1 - i}{2}\begin{pmatrix}
		1 & -i\\
		-i & 1 \end{pmatrix}
\end{equation}
and thus forms linear superpositions of the two polarizations. The resulting photocurrents on the detectors after the PBS are then given by
\begin{align}
	I_\mathrm{a} &\propto \left| E^\mathrm{(r)}_{||} + i E^\mathrm{(r)}_{\perp} \right|^2 \\
	I_\mathrm{b} &\propto \left| E^\mathrm{(r)}_{||} - i E^\mathrm{(r)}_{\perp} \right|^2.
\end{align}
Using the relation $|z_1 + i z_2|^2 = |z_1|^2 + |z_2|^2 + 2 \mathrm{Im}(z_1 z_2^*)$ it is easy to show that
\begin{equation}
	|z_1 + i z_2|^2 - |z_1 - i z_2|^2 = 4 \mathrm{Im}(z_1 z_2^*)
\end{equation}
Accordingly, the classic HC error signal is given by
\begin{align}
	V^\mathrm{HC}_\mathrm{error} &\propto I_\mathrm{a} - I_\mathrm{a} \\
								 &\propto 4 \mathrm{Im}( E^\mathrm{(r)}_{||}  (E^\mathrm{(r)}_{\perp})^*) \\
	                             &\propto E^\mathrm{(r)}_{\perp} \mathrm{Im}( E^\mathrm{(r)}_{||}) \\
	                             &\propto E_0^2 \sqrt{R_1} \sin(\theta) \cos(\theta) \mathrm{Im}( F^\mathrm{cav}_{||}(\Delta \omega) ) \\ \label{eq:app:V_err_HC}
	                             &\propto \mathrm{Im}( F^\mathrm{cav}_{||}(\Delta \omega) )
\end{align}
where we first simplified the expression realizing that $E^\mathrm{(r)}_{\perp}$ is purely real and later inserted the expressions from above. This shows that the classic HC method extracts the imaginary part of the detuning-dependent complex reflection coefficient $F^\mathrm{cav}_{||}(\Delta \omega)$ of the polarization probing into the cavity, as used in the main text.

\subsubsection{H\"ansch-Couillaud for injected laser diodes}\label{sec:app:theory:INJ}
In case of the adapted HC for laser diode injection, the seed light is again described by
\begin{equation}
	E^\mathrm{(i)} = E_0 \left( \cos(\theta)\hat{H} + \sin(\theta)\hat{V} \right). \label{eq:app:E_i_INJ}
\end{equation}
Only the horizontal component injects the laser diode leading to an allowable detuning range $\Delta\omega \propto E_0  \cos(\theta)$. The vertical component is reflected at the entrance facet of the laser diode with reflectivity $R_1$ and serves as the reference beam. The resulting back-reflection is given by
\begin{align}
	E^\mathrm{(r)} &= E_\mathrm{ILD} F^\mathrm{ILD}_{||} \hat{H} + E_0 \sin(\theta) F^\mathrm{ILD}_{\perp} \hat{V} \label{eq:app:E_r_INJ}
\end{align}
where $E_\mathrm{ILD}$ denotes the strength of the electric field emitted by the ILD, $F^\mathrm{ILD}_{||} = e^{i \phi(\Delta\omega)}$ from (Eq.~\ref{eq:F_inj}), and $F^\mathrm{ILD}_{\perp} = \sqrt{R_1}$. We note that strictly speaking $F^\mathrm{ILD}_{||}$ is not a reflection coefficient because the amplitude of the back-reflection is primarily determined by the diode drive current and is thus nearly independent of the amplitude $E_0 \cos(\theta)$ of the input light. However, this does not alter the argument in any meaningful way since primarily the phase shift of $F^\mathrm{ILD}_{||}$ is relevant, rather than the amplitude.\par

A polarization-dependent pick-off with reflectivities of $R^\mathrm{po}_{||}$ and $R^\mathrm{po}_{\perp}$ is chosen such that it roughly equalizes the field strengths of the two polarizations
\begin{align}
	E_\mathrm{ILD} \sqrt{R^\mathrm{po}_{||}} \approx E_0 \sin(\theta) \sqrt{R_1} \sqrt{R^\mathrm{po}_{\perp} } 
\end{align}
to avoid saturation of the amplified PDs.

The light going into the detector branch is then given by
\begin{align}
	E^\mathrm{(r,po)} = &E_\mathrm{ILD} \sqrt{R^\mathrm{po}_{||} } e^{i \phi(\Delta\omega)} \hat{H} + \notag \\
	                    &E_0 \sin(\theta)  \sqrt{R_1} \sqrt{R^\mathrm{po}_{\perp}}   \hat{V} \\ \label{eq:app:E_r_pickoff}
	              \equiv& E^\mathrm{(r)}_{||} \hat{H} + E^\mathrm{(r)}_{\perp} \hat{V}
\end{align}
With the usual detector arrangement based on a quarter-wave plate at 45$^\circ$ followed by a PBS and two PDs the error signal would again be
\begin{align}
	V^\mathrm{ILD}_\mathrm{error} &\propto E^\mathrm{(r)}_{\perp} \mathrm{Im}( E^\mathrm{(r)}_{||}) \\
	&\propto E_\mathrm{ILD}  E_0 \sin(\theta)   \sqrt{R_1 R^\mathrm{po}_{\perp} R^\mathrm{po}_{||} }  \mathrm{Im}( e^{i \phi(\Delta\omega) } ) \\
	&\propto \sin( \phi(\Delta\omega) ). \label{eq:app:V_error_inj_first}
\end{align}
Recalling that 
\begin{align}
\phi(\Delta\omega) = \sin^{-1} \left(\frac{\Delta\omega}{ \Delta\omega_\mathrm{min}  } \right) - \tan^{-1}(\alpha),
\end{align}
the only thing preventing a perfectly linear error signal is the term $\tan^{-1}(\alpha)$. It is possible to compensate for it, by introducing a relative phase shift $e^{i\gamma_\mathrm{rot}}$ between the horizontal and vertical polarizations, which rotates the coordinate system as previously illustrated in FIG.~\ref{fig:HC_theory}(e).\par

In practice, the phase shift $\gamma_\mathrm{rot}$ can be adjusted by rotating the quarter-wave plate before the PBS. However, an additional $\lambda/2$ waveplate is then necessary to restore the amplitude balance between the polarization required to generate an error signal without a DC offset, resulting in
\begin{align}
	V^\mathrm{ILD}_\mathrm{error} &\propto E_\mathrm{ILD}  E_0 \sin(\theta) \sqrt{ R_1 R^\mathrm{po}_{\perp} R^\mathrm{po}_{||} } \frac{\Delta\omega}{ \Delta\omega_\mathrm{min} }
\end{align}
shown in FIG.~\ref{fig:HC_theory}(f). In practice, $\gamma_\mathrm{rot}$ will also include additional relative phase shifts between the two polarizations, e.g. due to the optical pick-off near Brewster’s angle. The shape of the error signal depends strongly on the settings of the quarter-wave plate and the half-wave plate before the PBS, requiring careful alignment. For this, we have found the model involving projections in the complex plane to be a useful guide.

\end{document}